# Devices for high precision x-ray beam intensity monitoring on BSRF

LI Hua-Peng(李华鹏)[1, *] TANG Kun(唐坤)[1] ZHAO Yi-Dong(赵屹东)[1] ZHENG-Lei(郑雷)1 LIU Shu-Hu(刘树虎)[1] ZHAO Xiao-Liang(赵晓亮)[1] ZHAO Ya-Shuai(赵亚帅)[1]

1 Beijing Synchrotron Radiation Facility, Institute of High Energy Physics, Chinese Academy of Sciences, Beijing 100049, China

* hplee@ihep.ac.cn；

## Abstract:

Synchrotron radiation with the characteristic of high brilliance, high level of polarization, high collimation, low emittance and wide tunability in energy has been used as a standard source in metrology(1, 2). For a decade, lots of calibration work have been done on 4B7A in Beijing Synchrotron Radiation Facility (BSRF) (3, 4). For the calibration process, a high-precision online monitor is indispensable. To control the uncertainty under 0.1%, we studied different sizes parallel ion chambers with rare-gas and used different collecting methods to monitor the x-ray intensity of the beamline. Two methods to collect the signal of the ion chambers: reading the current directly with electrometer or signal amplification to collect the counts were compared.

Supported by National Natural Science Foundation of China (No. 11375227).

## 1. Introduction

With the rapid development of high technology and interdisciplinary subject，X-ray from synchrotron radiation has acquired growing importance in basic research and applications such as: metrology，biological science, microelectronics, materials science, surface science, astrophysics, laser plasma diagnostic technique and other applications. Meanwhile most of the researches request the absolute intensity of the incident X-rays beam, especially in metrology and laser plasma diagnostic technique(5). And lots of work had been done to meet these demands around the word. For example, the PTB has been strongly engaged in the field of metrology using synchrotron radiation in the past thirty years(6). The beamline 4B7A of BSRF also had calibrated lots of optical devices for years and established systematic calibration method, with the transfer detector standard which calibrated at the PTB(7). Now a cryogenic electrical-substitution radiometers (ESR) as Primary detector is under Development.

However, our source, BSRF, is the first generation synchrotron radiation facility and the e current decays from 250mA to 180mA, and the ESR and the detector to be calibrated can’t be in the beamline simultaneously. Therefore high precision online monitoring is necessary during the calibration work. The device will be put in the beamline between the last optical or beam-shaping element and the sample. So the device should fulfill two requirements: steady to meet high precision and transmission with low attenuation. Thus we developed low pressure rare-gas ion chamber to monitor the beam intensity.

## 2. Devices and experiment methods

Except the source, online monitoring system constituted by two components: signal generate system and signal readout system.

2.1 Devices for signal generate

We developed two ion chambers to study the characteristics of online monitoring. In order to reduce ion recombination, we designed the two parallel plate ion chambers (Figure 1) with different size of sensitive areas. To get enough high signal with low recombination loss, the collecting electrode of the ion chamber-A is 30 cm long while that of the ion chamber-B is 100 cm long. Each of them with detached collecting electrodes. Also, protecting electrode is necessary to keep the electric field in the sensitive area as uniformly as possible. All the electrodes are made of copper, isolated by PTFE. In order to obtain high precision signal, different energy with different work gas pressure and bias voltage will be studied.

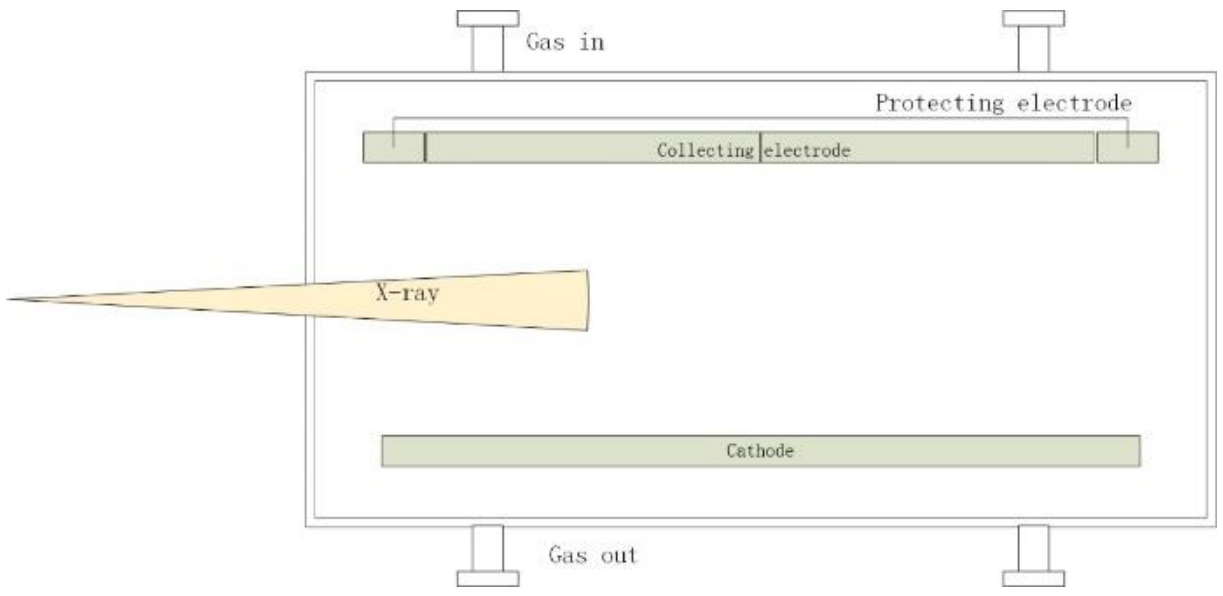

Figure 1 Sketch map of the ion chambers. The large one and the small one have the similar structure with different sizes.

2.2 Methods to obtain the signal

When X-ray pass through the ion chamber, part of the photons will be absorbed and signal will be generated:

$$N_{abs} = N_{in} - N_{out} = N_{in}\left(1 - e^{-\frac{L\sigma P}{KT}}\right), \qquad (1)$$

$N_{abs}$ is the absorbed photons, L is the absorb length of the sensitive area, $\sigma$ is the cross section and P is the pressure of the work gas. The absorbed x-rays $N_{abs}$ produce electron and ions, with biasing voltage the ions or electron will be collected. We developed two methods to record the signal: current mode and count mode (see Figure 2). With the current mode, we just use a Keithley 6517B electrometer to measure the current. With the count mode, we use an amplifier and V/F convert and 974A counter to get counts with certain integration time.

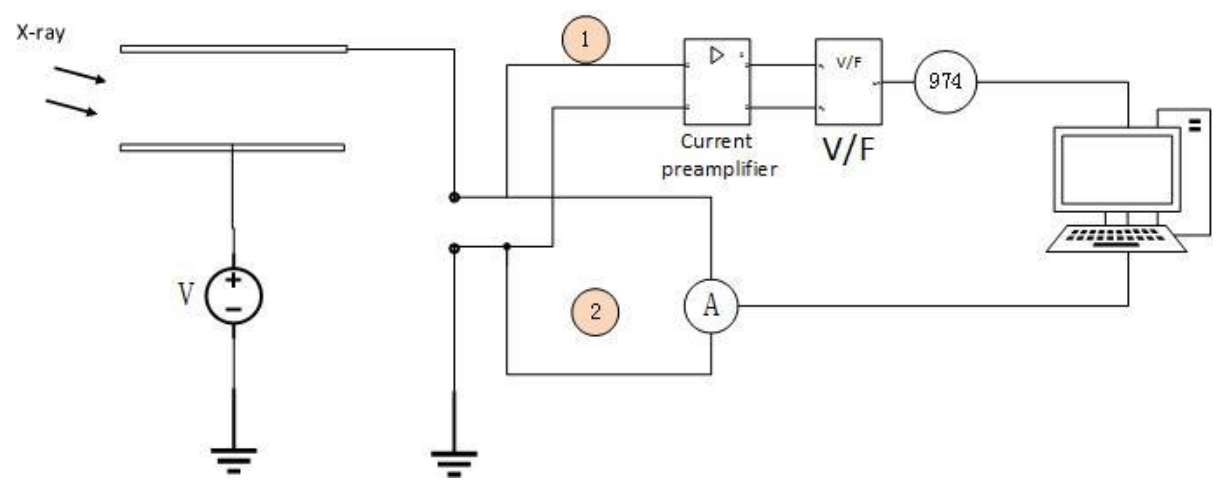

Figure 2 Two methods to obtain the signal: amplifier and V/F convert and 974A counter to get counts (1); directly read the current (2).

2.3 signal strength

The statistical error of the measured current of the ionization chamber should be low enough to obtain the high precision monitoring. As for ion chamber, the statistical error $\sigma_s$ follows Poisson statistics(8), and related to the collected charge quantity in unit time:

$$\sigma_s = \frac{\sqrt{N_{ion}}}{N_{ion}} \text{ , (2)}$$

In order to get high precision uncertainty less than 0.1%, the statistical should be less than 0.01%, thus the $N_{ion}$ should large than 10^8/s means that current large than 10 pA.

$$N_{ion} = N_{abs}\frac{E}{w}, \text{ (3)}$$

E is the photon energy, and W is the average energy necessary for the creation of one electron ion pair. For 3keV photon pass through Kr, with the W=24±0.7ev(9)， the $N_{abs}$should large than 10^6. And our source, 4B7A in BSRF, has the flux as large as 1E10 photon/s. thus the Poisson fluctuations can be neglected.

# 3. Experiments

The experiments were performed at the 4B7A beam line on BSRF. The energy of the electrons is 2.5GeV in the storage ring, where the beam current is 180 mA–250 mA. The magnetic field of the bending magnet is 0.808 T with critical energy of 3358.6 eV. The size of the source is about 1.5 mm (H) × 0.4 mm (V). The vertical divergence of the source at the critical energy is about 0.28 mrad. The maximum horizontal acceptance angle is 5 mrad which is defined by the apertures in the front-end section. This beamline is equipped with a Double-Crystal Monochromator (DCM, made by KOHZU, Japan) with fixed incident and exit beam height. The first crystal in DCM is cooled by water with temperature of 20° Celsius to keep thermal stability. Usually used crystals are Si(111) and InSb(111). The corresponding energy range is from 1.75 to 3.5 keV while using InSb(111), and from 2.1 to 6.0 keV for Si(111)(10).

In order to obtain steady and accurate beam intensity, the ion chamber should be filled with proper gas pressure and implemented with appropriate bias voltage. Then the saturation curve of different pressures will be studied, and the attenuation of the chamber will be examined in case of the weak beam intensity(11).

3.1 Ion chamber saturation curve

This experiments were carried out with a monochromatic beam at 3keV. We used Kr as working gas, Figure 3 shows the ionization chamber plateaus of different pressure of Kr with different ion chambers. As shown in the picture, with different pressures, the saturation seems in the same area. With lower pressure or smaller sensitive area, the plateau curve reaches saturation more quickly with lower voltage. With the same pressure of the two ion chambers, means the same recombination mode (12-15), chamber B have a saturation current one order of magnitude larger than chamber A. While the signal magnitude of chamber A is more than sufficient for the demand. So chamber A has been used.

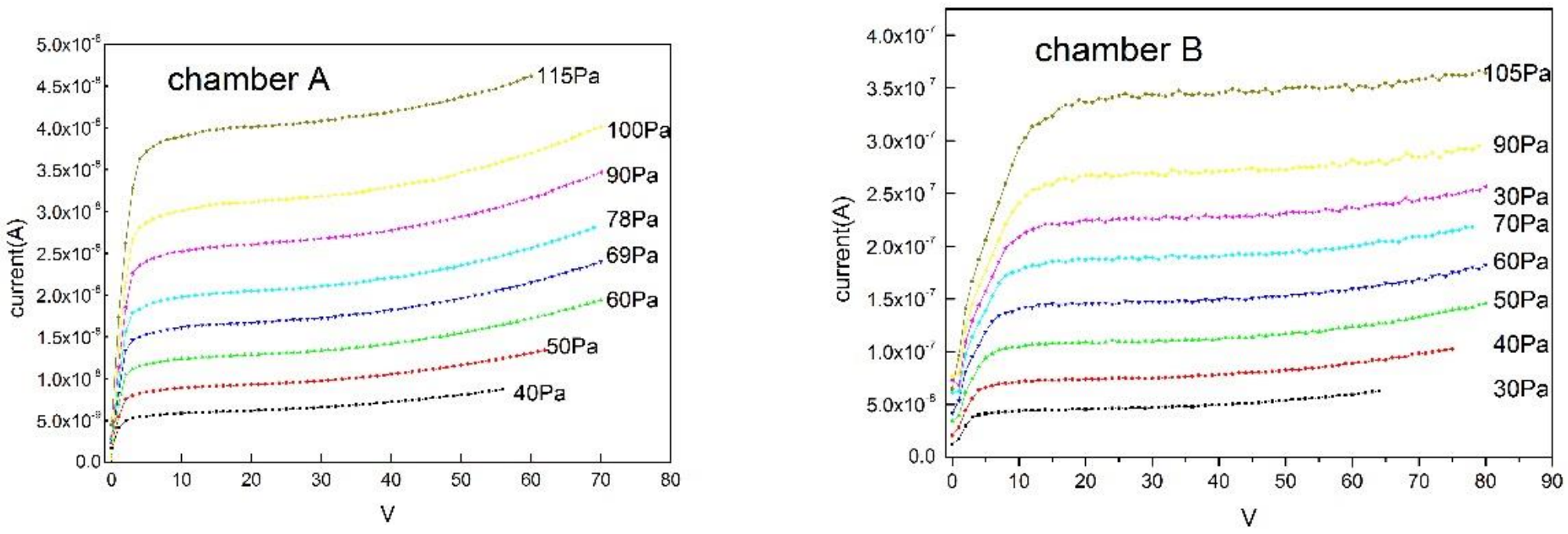

Figure 3 Left is chamber A with different pressure of Kr, while the right is chamber B.

## 3.2 Attenuation of the ion chamber

We also studied the transmission of the beam with different pressure of Kr with small ion chamber. We used two Si photodiode (AXUV100) which has been calibrated at PTB ahead and behind the small ion chamber. The result has shown in figure 4, and has been compared with the CXRO database(16). With the result, we can choose proper pressure while the transmission is acceptable.

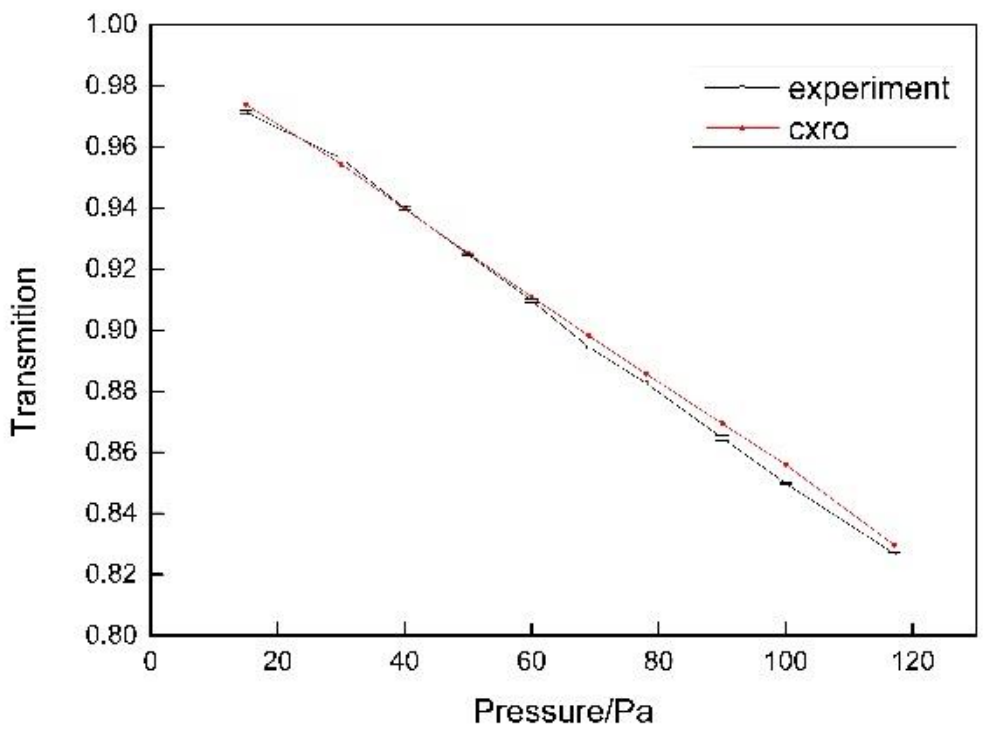

Figure 4 The black open block is the experiment result in 3 Kev with error bar, the error bars are smaller than the sizes of the symbols, and the solid block in red is the result from CXRO.

## 3.3 Linearity

For a proper performance of the monitor the detector signal must depend linearly on the incident photon flux. As shown in the figure 3, the plateaus of chamber A, with different pressure of Kr, the saturation voltage seems at the same area 15v-40v. Thus, we filled the chamber A with 50pa Kr, and the bias voltage is 20V. With the change of different filters, we got different signals of the ion chamber. And we assume that the flux without filter as 100%. Each measurement of different flux are normalized with a beam monitor. The result was shown in the Figure 5. And good linearity has been received.

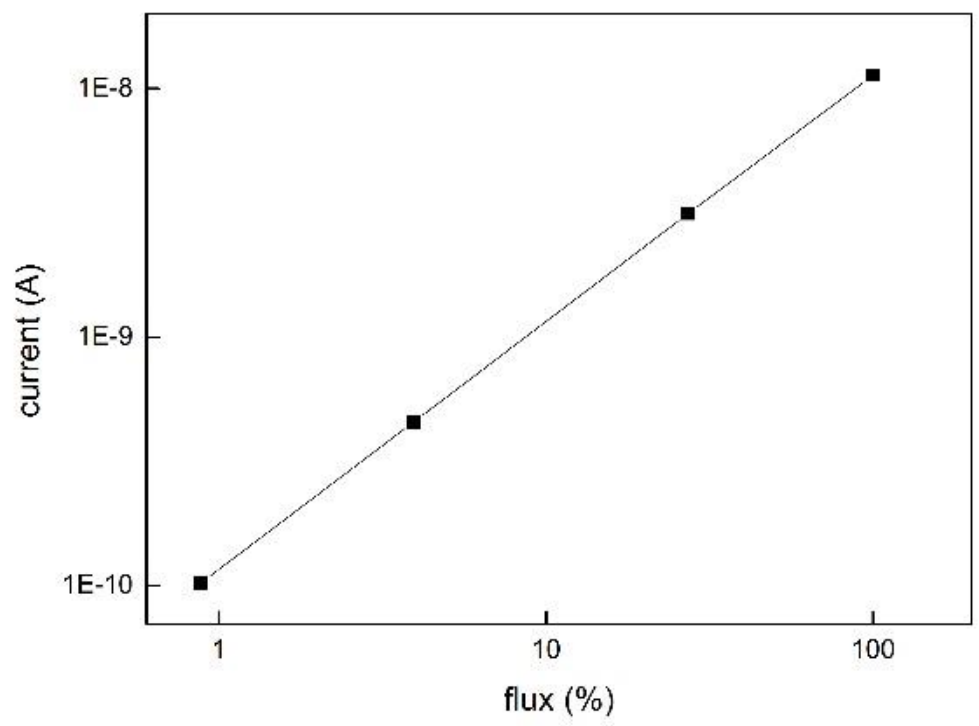

Figure 5 Dose rates were relatively varied: 0.88 %, 3.93 %, 27.15 %, and 100 %. The current of the ion chamber increasing. And the result shows that the Adj. R-Square is 0.99994.

# 4. Results and analysis

The signal of the chamber can be estimated by calculate the number of created electron = ion pairs, which depends on the constituent, pressure and temperature of the working gas, the active length of the ion chamber, and the energy and flux of the incoming photons. As shown in formula (1), the change of the working gas pressure, temperature, energy and flux of the incoming photons will affect the signal. We expect to measurement the flux with low uncertainty. Thus the uncertainty caused by the gas pressure, temperature and energy of the photons should be concerned. Besides the device to obtain the signal will bring electronic noise. In addition to the Poisson fluctuations, there are other additive noise sources in the system, e.g., electronics noise and uncertainty caused by P, T and E. Therefore the total uncertainty can be

$\sigma_{total}{}^2 = \sigma_s{}^2 + \sigma_{elec}{}^2 + \sigma_P{}^2 + \sigma_T{}^2 + \sigma_E{}^2$, (4)

As shown before, the Poisson fluctuations $\sigma_s$, according to our photon flux, will be less than 0.01%. The uncertainty of the pressure will be

$$\sigma_p = \frac{N_{abs,\ \mathrm{p+\Delta p}} - N_{abs,\ \mathrm{p}}}{N_{abs,\ \mathrm{p}}} = \frac{\frac{L\sigma P}{KT} e^{-\frac{L\sigma P}{KT}}}{1-e^{-\frac{L\sigma P}{KT}}} \frac{\Delta P}{P}, \quad (5)$$

Similarly the temperature can be

$$\sigma_T = \frac{N_{abs,\ \mathrm{T+\Delta T}} - N_{abs,\ \mathrm{T}}}{N_{abs,\ \mathrm{T}}} = \frac{-\frac{L\sigma P}{KT} e^{-\frac{L\sigma P}{KT}}}{1-e^{-\frac{L\sigma P}{KT}}} \frac{\Delta T}{T}, \quad (6)$$

As shown in the formula (5) and (6), the uncertainty of P and T can be neglected, for we have tested the vacuum performance of the chamber and the experiment is performed in a constant temperature environment. As for the energy, the resolution of our beam line is 0.2ev, thus $\sigma_E$ can be 6.7E-5. Therefore the most important influence can be electron noise.

According to result before, we performed the large ion chamber with 50pa Kr and 30V to study the electron noise, the energy was 2500eV. We put the large ion chamber in the beamline and collect the signal of the two collecting electrodes at the same time. Thus the energy, flux, pressure and temperature are all the same. Then uncertainty of the ratio of two electrode is mainly caused by the electron noises. We used two collecting methods separately for a time. The ratio was shown in the figure 6. As shown in the picture, the count mode (solid cubic) are less fluctuation than the current mode (red line). Statistics on this two result, the current uncertainty is 0.16%, and the count mode is 1.56E-4. As we can see, the count mode shows better preforms, for it integral for a time.

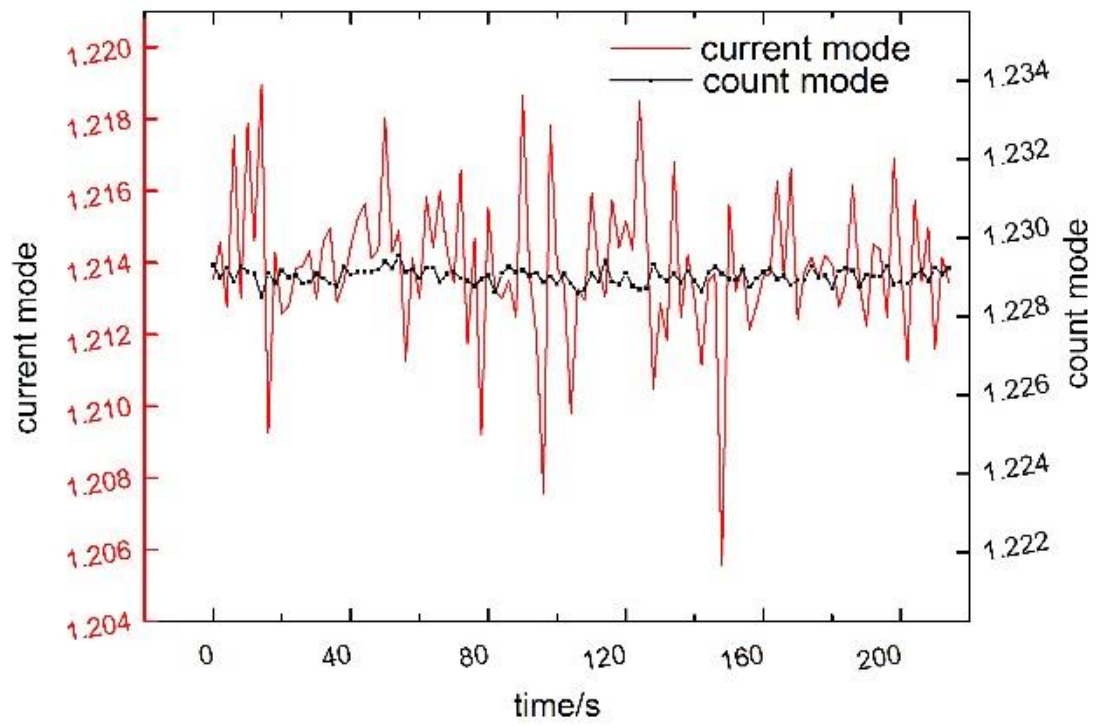

Figure 6 The line is the curve stands for the ratio got in current mode, while the solid cubic is obtained by the count mode. Statistics on this two result, the current uncertainty is 0.16%, and the count mode is 1.56E-4.

## 5. Conclusion

We built two sizes of rare-gas ion chambers for high precision on line monitoring on BSRF. The beam monitor is a parallel ionization chamber that working at alterable gas pressure, for one can use proper pressure for certain photon energy to obtain steady and high precision working state. It is designed to operate in the energy range of 2.1-5.5kev without a significant attenuation of the original beam. The linearity of the ionization chamber signal was shown and the Adj. R-Square is 0.99994. We also studied different signal collect method to depress the uncertainty, analysis all the factors that affect the monitoring performance to get steady and high precision on line monitoring system. Provide accurate state of the beam intensity for calibration and XAS work. We put the on line monitoring system in the beamline and it working good during the calibration and XAS work.

## 6. Acknowledgement

Thank you for all the staffs on the 4B7A beamline for their help during the measurements at the 4B7A beamline.

1. Richter M, Ulm G. Radiometry using synchrotron radiation at PTB. Journal of Electron Spectroscopy and Related Phenomena 101–103 (1999) 1013–1018. 1999.

2. Gottwald A, Klein R, Müller R, Richter M, Scholze F, Thornagel R, et al. Current capabilities at the Metrology Light Source. Metrologia. 2012;49(2):S146-S51.

3. Zhao YD, Cui MQ, Pei-Ping. Z, Gang. L, He-Sen. T, Ren-Jian. Z, et al. Cablibration of soft X-ray Detector for Astronomic Obserbation. HIGH ENERGY PHYSICS AND NUCLEAR PHYSICS. 2001;25(2):205-12.

4. Ke-xu S, Rong-oing Y, Shao-En J, Jia-Min Y, Tian-xuan H, Yan-Li C, et al. Calibration of Soft X-Ray Detection with Synchrotron Radiation. HIGH ENERGY PHYSICS AND NUCLEAR PHYSICS. 2004;28(2):205-9.

5. D Y, Z W, X J, Y L, X P, T Z, et al. Accurate and efficient characterization of streak camera using etalon and fitting method with constraints. Review of Scientific Instruments. 2011;82(11):113501 - -5.

6. Beckhoff B, Gottwald A, Klein R, Krumrey M, Müller R, Richter M, et al. A Quarter-Century Of Metrology Using Synchrotron Radiation By Ptb In Berlin. Physica Status Solidi. 2009;246(7):1415–34.

7. Rabus H, Scholze F, Thornagel R, Ulm G. Detector calibration at the PTB radiometry laboratory at BESSY. 1996

8. Kocsis M, Somogyi A. Miniature ionization chamber detector developed for X-ray microprobe measurements. J Synchrotron Rad (2003) 10,187-190. 2003.

9. Saito N, Suzuki IH. Photon w Value for Krypton in the M-Shell Transition Region. 2001.

10. Zheng L, Zhao YD, Tang K, Ma CY, Hong CH, Han Y, et al. A new experiment station on beamline 4B7A at Beijing Synchrotron Radiation Facility. Spectrochimica Acta Part B: Atomic Spectroscopy. 2014;101:1-5.

11. Ahmed SN, Besch H-J, Walenta AH, Pavel N, Schenk W. High-precision ionization chamber for relative intensity monitoring of synchrotron radiation. Nuclear Instruments and Methods in Physics Research A 449 (2000) 248}253. 1999.

12. GREENING JR, F.INsT.P. Saturation Characteristics of Parallel-Plate Ionization Chambers. 1963.

13. KIATEL MT. An experimental study of ion recombination in parallel-plate free-air ionization chambes. 1967;PHYS. MED. BIOL., 1967, VOL. 12, NO. 4, :555-63.

14. Bohm J. Saturation corrections for plane-parallel ionisation chambers. Phys Med Biol. 1976;21(5):754.

15. Nariyama N. Ion recombination in parallel-plate free-air ionization chambers for synchrotron radiation. Phys Med Biol. 2006;51(20):5199-209.

16. http://www.cxro.lbl.gov/. The Center for X-Ray Optics.